# Harnessing Native Chromium Oxidation for Giant Orbital Torque and Field-Free Magnetization Switching in NiFe/Cr Bilayers

Chirag Kalouni[a,1,2], Abhishek Kumar[a,1], Manish Kumar Mohanta[3,4], Kanupriya[1], Athira Ravindran K[1], Damanpreet Kaur[5], Preet Kamal[1], Puru Jena[4], and Debangsu Roy[1*]

[1]Department of Physics, Indian Institute of Technology Ropar, Rupnagar 140001, Punjab, India
[2]School of Mechanical and materials Engineering, Indian Institute of Technology Mandi, Himachal Pradesh 175005, India

[3]Department of Physics, Indian Institute of Technology Bhubaneswar, Bhubaneswar 752050, Odisha, India
[4]Department of Physics, Virginia Commonwealth University, Richmond, VA 23284, USA
[5]Department of Physics and Materials Science and Engineering, Jaypee Institute of Information Technology Noida, Gautam Buddha Nagar, Uttar Pradesh 201309, India

[a]Equal Contribution.

[*]Correspondence: debangsu@iitrpr.ac.in

Orbital currents offer charge-to-spin conversion beyond the efficiency limit of conventional heavy-metal Spin Hall sources. However, harnessing them has so far required either thick orbital-Hall materials or additional heavy-metal conversion layers. Here, we show that the native oxide of chromium, typically regarded as parasitic, transforms a simple NiFe/Cr into a self-contained dual-channel orbital-current source without the need for any conversion layer. First-principles calculations predict a nearly threefold enhancement of the orbital Hall conductivity upon surface oxygenation, driven by Cr:3d–O:2p hybridization. Experimentally, naturally oxidized NiFe/Cr heterostructures exhibit a giant damping-like torque efficiency of $3.9 \times 10^6\, \Omega^{-1} m^{-1}$, exceeding Pt (Ta) by one (two) orders of magnitude. The torque depicts a non-monotonic Cr-thickness dependence which cannot be explained by a conventional model. We have developed a drift-diffusion model with an oxidation-gated interfacial source which quantitatively reproduces the data, revealing that the Cr/$CrO_x$ interface generates orbital currents over an order of magnitude stronger than the bulk orbital Hall channel with an orbital transport length of $\approx 4\, nm$. The enhanced torque enables field-free magnetization switching at $1.58 \times 10^{11}\, A\, m^{-2}$, outperforming heavy-metal and $CuO_x$ benchmarks. These results establish native oxidation as a scalable strategy for realizing efficient orbital-torque devices.

Keywords: Orbitronics, Field free switching, Chromium Oxide, OHE, OREE.

## 1| Introduction

Electrical control of magnetization through current-induced torques has emerged as the foundation for energy-efficient non-volatile memory and logic devices, with significantly faster writing than conventional magnetic-field-based systems[1]. In spin–orbit torque (SOT) devices, an in-plane charge current is converted into a transverse spin current which exerts a torque on an adjacent ferromagnet, enabling deterministic magnetization switching in the adjacent ferromagnet[2]. Here, the charge-to-spin conversion has traditionally relied on the spin Hall effect (SHE) and the interfacial Rashba–Edelstein effect in materials with strong spin-orbit coupling (SOC) for example in heavy metals like Pt, $\beta - Ta$ or $\beta - W$[3-7]. However, their performance is fundamentally constrained because strong SOC simultaneously reduces the spin diffusion length and is often accompanied by low electrical conductivity.

The orbital Hall effect (OHE) has recently emerged as a promising alternative capable of overcoming these limitations[8-9]. In the OHE, a charge current generates a transverse flow of orbital angular momentum rather than spin angular momentum. When injected into an adjacent ferromagnet, this orbital current is converted into spin through SOC, thereby producing an effective spin torque[10-11]. Since the generation of orbital current does not require strong SOC, the OHE can occur in a broad range of light transition metals (3d to 5d)[11-15]. Remarkably, several of these materials are predicted to possess orbital Hall conductivities substantially larger than the spin Hall conductivity of the best heavy-metal spin sources, making them attractive candidates for low-power orbitronics devices[16-19].

Despite its promise, directly utilizing the bulk OHE presents several challenges. First, orbital currents do not directly exert a torque and must be converted into spin currents through SOC inside the ferromagnet or an adjacent conversion layer[11, 20]. Consequently, the highest orbital torque efficiencies reported so far frequently employ a heavy-metal layer, typically Pt, as an orbital to spin converter[21, 22]. This ironically reintroduces the heavy metals that orbitronics initially aimed to replace. Second, the exact orbital diffusion length remains under debate, with reported values ranging from sub-nanometre to several tens of nanometres[11, 23]. Nevertheless, bulk OHE devices typically require relatively thick orbital source layers to maximize the generated torque, which ultimately results in higher operating currents. For instance, Chromium, has recently been demonstrated to generate orbital torques comparable to those of β-W in Cr/CoFeB/MgO heterostructures, but only when both the Cr layer and the ferromagnet are made sufficiently thick[24].

An alternative approach exploits the interfacial Orbital Rashba-Edelstein effect (OREE). Here, the orbital accumulation is generated directly at an inversion symmetry breaking interface instead of within the bulk [22, 25-26]. The most extensively investigated OREE platform to date is oxidized copper, in which the $Cu/CuO_x$ interface substantially enhances the damping-like torque efficiency and enables current-induced magnetization switching[25, 27-33]. However, there are fundamental constraints associated with $CuO_x$ based systems. Firstly, the uncertainty surrounding the microscopic origin of the observed torques in $CuO_x$ based systems. Metallic Cu itself is a very weak bulk orbital-Hall conductor, implying that the orbital response arises predominantly from the oxide interface, making Cu layer merely a passive transport channel [29-30]. Further, competing interfacial mechanisms such as interfacial orbital hybridization, and spin–vorticity coupling driven by the oxygen induced drift velocity gradient apart from OREE are predicted to be the origin for the interfacial torque[21, 32, 34-38]. This is further complicated by the naturally inhomogeneous oxidation of Cu, which produces mixed oxide phases, grain boundaries, and composition gradients that obscure the structure of the active interface[27]. The second limitation parallels that of bulk orbital Hall systems. Because Cu contributes negligibly, the interfacial orbital current still requires conversion through a heavy-metal layer, typically Pt, to generate efficient spin torque[21-22, 26, 34, 39]. These drawbacks naturally raise the question of whether a light metal possessing both a strong intrinsic orbital Hall conductivity and a stable native oxide could simultaneously provide bulk and interfacial orbital-current generation within a single material system.

Chromium satisfies these requirements remarkably well. First-principles calculations predict that body-centred cubic Cr[12] possesses a giant intrinsic orbital hall conductivity of approximately $8200\ (\hbar/e)\ (\Omega\ cm)^{-1}$ while exhibiting a comparatively small spin hall conductivity ($\sim -130(\hbar/\mathrm{e})\ (\Omega\ \mathrm{cm})^{-1}$). More recently, direct magneto-optical measurements have experimentally confirmed substantial current-induced orbital accumulation in single-crystalline Cr, independent of any adjacent ferromagnetic layer, establishing Cr as a genuine bulk orbital-current source[17, 19, 24, 40-41]. Simultaneously, Cr readily forms a thin, chemically stable, self-limiting $Cr_2O_3$ layer under ambient conditions, naturally creating an inversion-symmetry-breaking Cr/$CrO_x$ interface[42]. Such a structure, in principle, offers a unique opportunity to investigate the coexistence of bulk orbital current generation in metallic Cr and a possible interfacial Orbital Rashba–Edelstein effect at the Cr/$CrO_x$ interface. However, whether these two mechanisms can coexist and operate synergistically, remains an open question. If realized, such a dual-channel architecture could provide an exceptionally efficient source of orbital torque, eliminating the need for an auxiliary heavy-metal converter while enabling field-free magnetization switching at low current densities.

## 2| Results and Discussion

In order to validate the proposed dual-channel nature of Cr, we first performed a comprehensive theoretical investigation of the constituent systems of $Ni_3Fe$ and Cr and of their heterointerfaces with and without surface oxygenation, namely $Ni_3Fe$/Cr-O and $Ni_3Fe$/Cr. The electronic and orbital-transport properties of the individual constituents and of the heterointerfaces were systematically analysed. The two-layer (2L) $Ni_3Fe$ structure is constructed from the (111) surface of fcc $Ni_3Fe$, whereas the Cr layers are modelled from the (001) surface of bcc Cr, following the previous experimental reports[17]. The electronic band structures of the various Cr layers, computed with spin–orbit coupling included, are presented in Figure S1.1(a,b,c). Band structure calculations (Figure S1.1) confirm the metallic character of the Cr layers, evidenced by multiple bands crossing at the Fermi level. The orbital Hall conductivity (OHC) of Cr films with different layer thickness was computed, revealing a strong dependence on the number of layers, as shown in Figure S1.1(a,b,c).

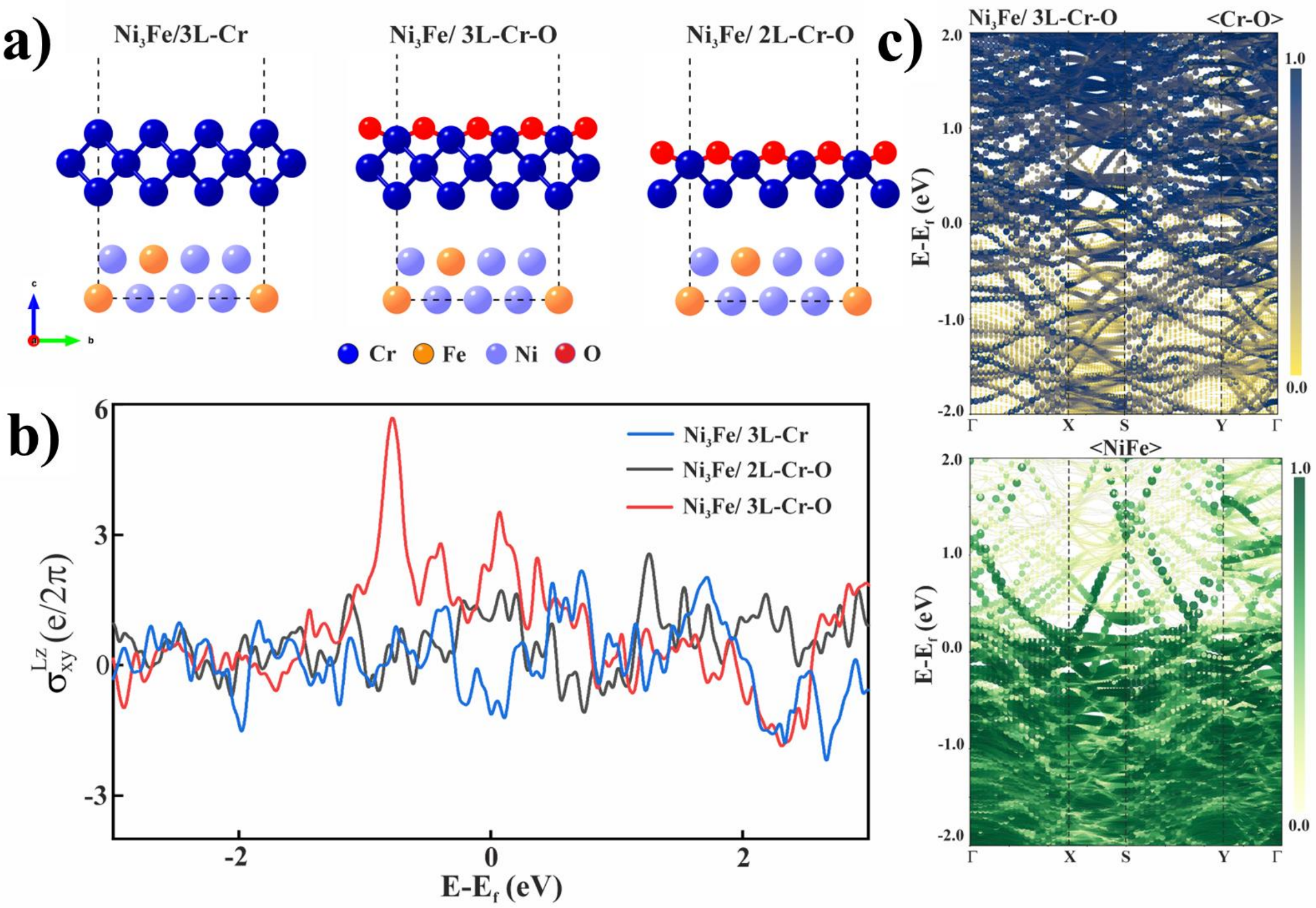

**Figure 1 |** (a) Geometrical view of different heterointerfaces studied, (b) corresponding orbital Hall conductivities as a function of energy, (c) layer projected electronic band structure of $Ni_3Fe$/3L-Cr-O.

A $Ni_3Fe$/3L-Cr heterointerface is constructed using a supercell approach to minimize the interfacial strain for theoretical investigation[43-44]. The geometrical view is shown in Figure 1a. Since our aim is to exploit oxidized Cr as the orbital source, the top Cr surface was correspondingly oxygenated in the model, as shown in Figure 1a. The calculated OHC of the heterointerface is presented in Figure 1b. Upon oxidation, the OHC of the $Ni_3Fe$/3L-Cr heterostructure is markedly enhanced, increasing from 2.12 to 5.65 (e/2π); an approximate threefold increase. The layer-projected band structure and density of states (DOS) in Figure 1c and Figure S1.2 respectively, reveal that the conduction bands near the Fermi level originate predominantly from the Cr-O layers. Further, a clear Cr-layer dependence can also be observed, with the OHC of $Ni_3Fe$/2L-Cr-O is 2.54 $\frac{e}{2\pi}$. These results demonstrate that the OHC is governed by both the Cr-layer thickness and the oxidation of the top Cr layer, with oxidation producing a pronounced enhancement of the orbital Hall response (Figure 1b). Figure 2a gives the geometrical view of 3L-Cr-O heterostructure. To elucidate the microscopic origin of this enhancement, the atom- and orbital-projected band structures of the 3L-oxidized Cr are shown in Figure 2b,c. The bands near the Fermi level in the oxidized Cr surface are predominantly derived from Cr: *d* orbitals. The DOS plot shown in Figure 2c further suggests a minor contribution from the O: *p* states arising through hybridization, while the dominant conduction pathways are governed by Cr: *d* states, with the Cr:3d–O:2p hybridization near $E_F$ providing a natural microscopic origin for the enhanced OHC**.** Moreover, these results suggest that incorporating an oxidized Cr interface with $Ni_3Fe$ can substantially enhance the overall OHC, offering a route to improved device performance.

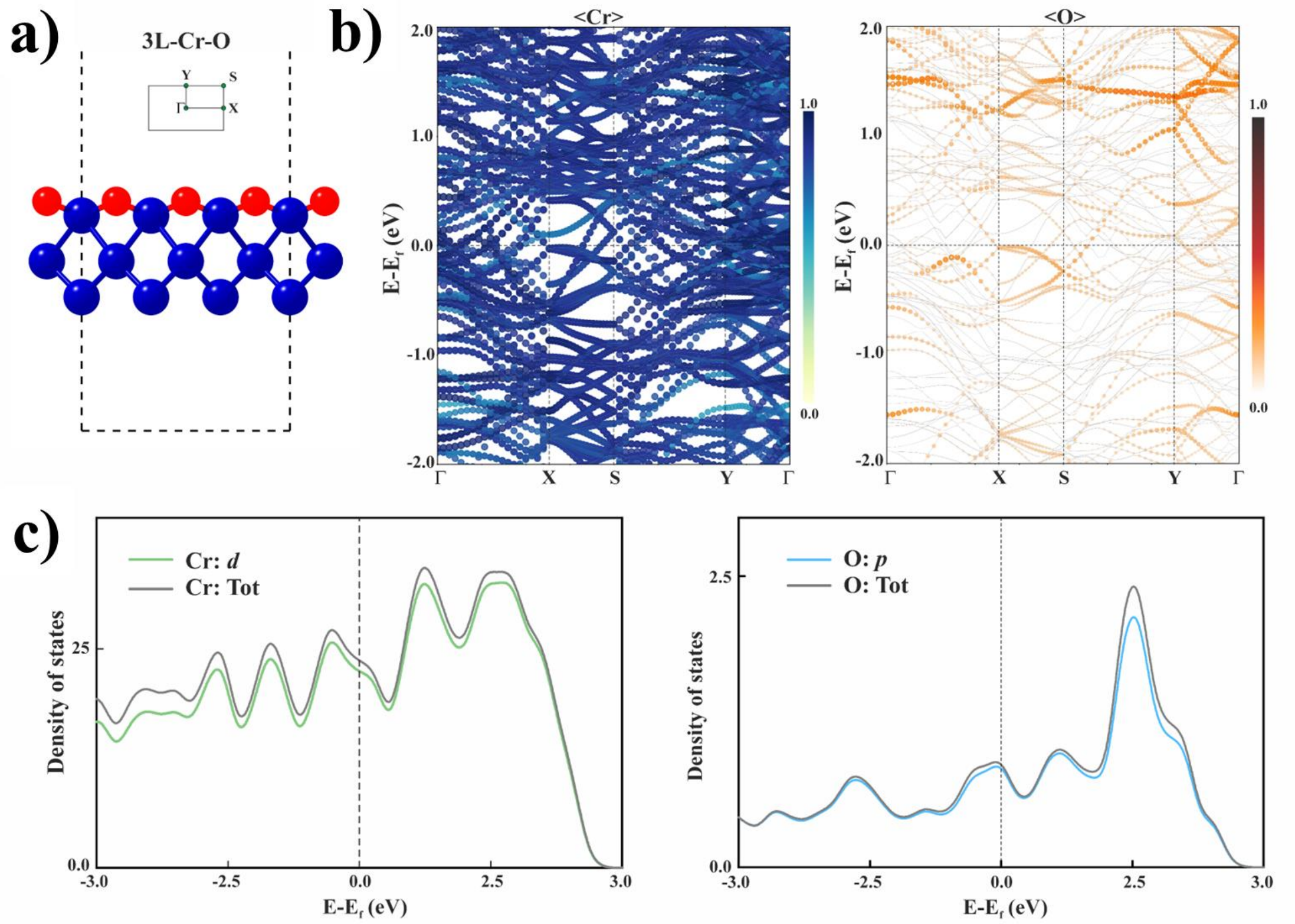

**Figure 2 |** (a) Geometrical structure of 3L-CrO, (b) corresponding atom (Cr/O) projected band structure, and (c) orbital projected density of states (DOS).

Motivated by these theoretical predictions, we fabricated FM/LM−LM−oxide heterostructures of the form Si/$SiO_2$/Ta(2)/NiFe(5)/Cr*(t), where the numbers in parentheses denote the layer thickness in nanometres. In this work, the Cr* thickness was varied from 4 to 25 nm.

Now to characterize the native oxidation of the Cr layer and establish the effective composition of the heterostructure, we performed elemental depth profiling by X-ray photoelectron spectroscopy (XPS) combined with $Ar^+$-ion etching on NiFe(5)/Cr*(6.5) and NiFe(5)/Cr*(11) stacks (Figure 3a). The spectra acquired at the surface (zero etching) show Cr together with oxygen. Upon successive etching, the oxygen signal persists, confirming that the oxidation extends beneath the top surface into the Cr layer (Figure S3(a)). The depth-resolved Cr 2p spectra reveal the coexistence of oxidized and metallic chromium, with metallic Cr appearing progressively beneath the oxide (Figure 3b and S3(b)). The gradual evolution of the Cr 2p peak position and intensity with depth indicates that the oxidation is not abrupt but instead forms a vertically graded self-limiting oxide layer. Notably, the oxidation level decreases from the top surface towards the bottom of the Cr film. The dominant oxide species is identified as $Cr_2O_3$, and the associated oxygen gradient extends approximately 3 nm from the top surface.

The O 1s spectra (Figure S3(c)), in conjunction with the Cr 2p analysis, further confirm the formation of this graded oxide. The metallic oxide peak around 530 eV corresponds to the dominant $Cr_2O_3$ oxide at the initial etch rates. As the etching depth increases, the O 1s signal decreases, marking the transition from oxidized Cr to metallic Cr. At higher etch levels, the O 1s peaks emerge again, signifying the contribution of oxide at the bottom interface, specifically from the underlying $SiO_2$ substrate and the oxidized Ta seed layer $TaO_x$. The magnetic NiFe layer, however remains unoxidized.

These measurements establish that the native oxidation produces a self-limiting Cr/$CrO_x$ heterostructure on top of NiFe, in which a graded $Cr_2O_3$ surface oxide coexists with an underlying metallic Cr layer. Since the NiFe layer itself remains unoxidized and the Ta seed oxidizes only from the substrate side, the effective system under study is NiFe/Cr*— a single light metal layer that simultaneously hosts a metallic Cr bulk-orbital channel and an inversion symmetry breaking Cr/$CrO_x$ interface. Hereafter, Cr* denotes the combined thickness of the metallic Cr and its graded native oxide.

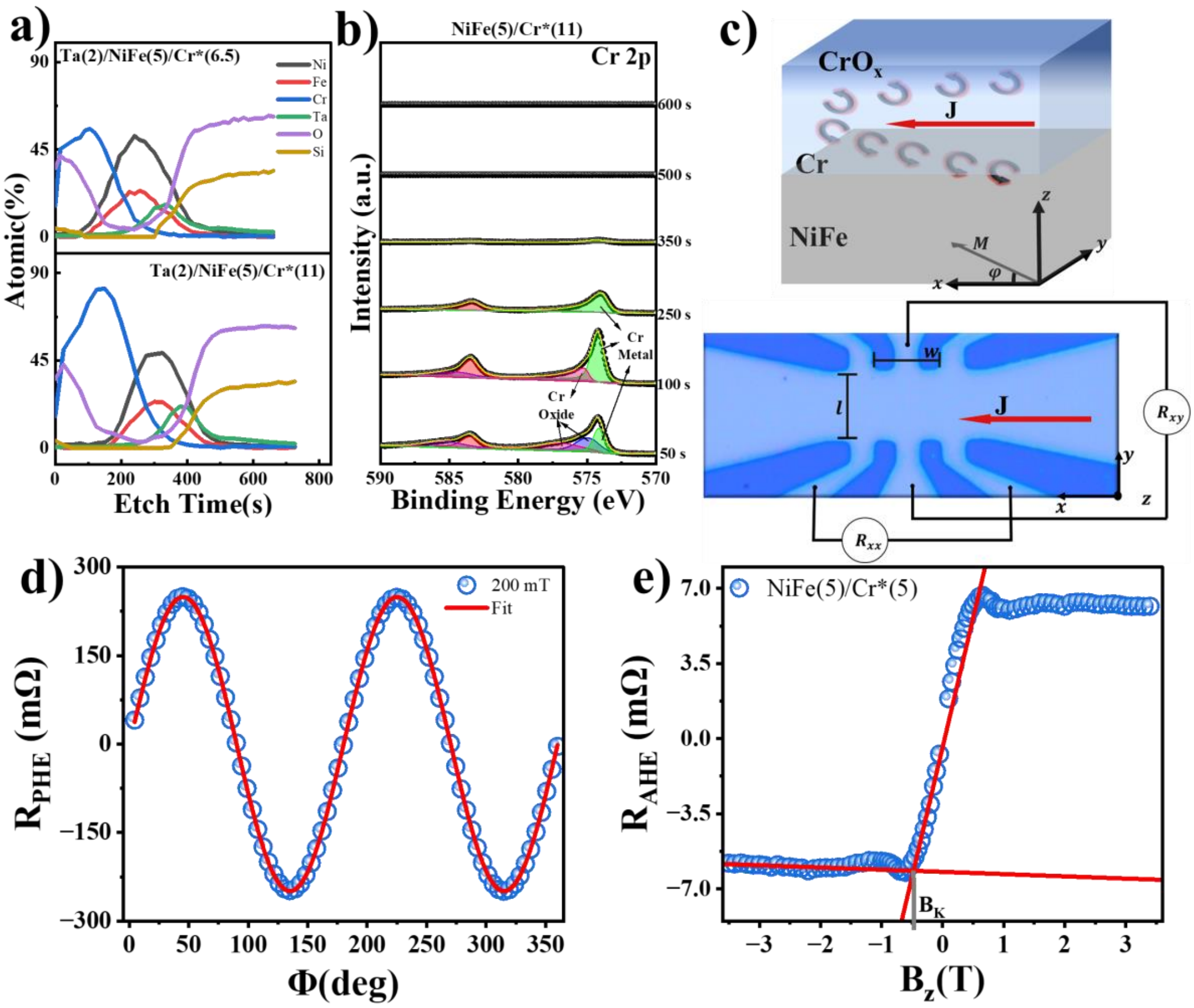

**Figure 3 |** (a) Atomic concentration of elements of the heterostructure. The formation of natural oxide is evident with the significant O contribution over Cr. (b) Cr 2p spectra of NiFe(5)/Cr*(11) stack. (c) Illustrative schematic for NiFe/Cr* heterostructures (top) and optical image of the hall bar (bottom). (d) Planar Hall resistance depicting the in-plane magnetic anisotropy at 200 mT for NiFe(5)/Cr*(5) sample at 3mA. (e) Out of Plane (OOP) sweep showing the anomalous hall resistance.

Next, to quantify the orbital torque generated by the Cr* layer, we performed angular-dependent harmonic Hall measurements on NiFe(5)/Cr*(t) devices. The schematic of the heterostructure and the corresponding Hall-bar device is shown in Figure 3c. We first characterized the fundamental magnetic properties of the NiFe(5)/Cr*(5) device via first-harmonic Hall resistance measurements. The in-plane angular dependence of $R_{xy}^{1\omega}$ exhibits a characteristic $sin\ 2\varphi$ behaviour (Figure 3d), confirming the in-plane magnetic anisotropy of the NiFe layer and yielding a planar Hall resistance,$R_{PHE} = (249.14 \pm 0.14)\ m\Omega$. Here, $\varphi$ is the angle made by the in plane $B_{ext}$ to the current direction. Subsequently, an out-of-plane magnetic field sweep was utilized to extract the anomalous Hall resistance (Figure 3e), $R_{AHE} = 6.3 \pm 0.1\ m\Omega$ and the effective perpendicular anisotropy field $B_K = 0.46 \pm 0.001$ T. These parameters are essential for the accurate quantification of the current-induced effective fields (See S4).

Next, second-harmonic Hall resistance $R_{xy}^{2\omega}$ measurements were performed under an external in-plane magnetic field $B_{ext}$ to extract the torque for various applied current amplitudes. The angular dependence of $R_{xy}^{2\omega}$ can be described by the following relation[45-46]:

$$R_{xy}^{2\omega}(\varphi) = R_{AHE}^{1\omega}.\left(\frac{B_{DL}}{B_{eff}} + R_{\nabla T}\right).\cos\varphi + 2R_{PHE}^{1\omega}.\left(\frac{B_{FL+Oe}}{B_{ext}}\right).(\,2cos^3\varphi - cos\varphi) \quad (1)$$

where $B_{eff}$ $(B_{ext} + B_K)$ is the summation of applied field and the effective perpendicular anisotropy field, $B_{DL}$ is the damping-like effective field, $B_{FL+Oe}$ encompasses both the field-like torque and the Oersted field contributions, and $R_{\nabla T}$ accounts for thermoelectric (Spin See beck and anomalous Nernst) contributions due to out of plane temperature gradient. Figure 4a shows $R_{xy}^{2\omega}(\varphi)$ for NiFe(5)/Cr*(5) at $B_{ext}$ = 250 mT. The $cos\ \varphi$ and $(\,2cos^3\varphi - cos\varphi)$ components were isolated by symmetrizing and anti-symmetrizing the raw data and fitted to Equation (1). By fitting the angular-dependent $R_{xy}^{2\omega}$ curves measured at different $B_{ext}$ fields using Equation (1), we extract the coefficient of the $cos\ \varphi$ and $(\,2cos^3\varphi - cos\varphi)$ terms. Subsequently, the linear fits of $R_{xy_\cos\varphi}^{2\omega}\ vs\ 1/B_{eff}$ and $R_{xy_2cos^3\varphi - cos\varphi}^{2\omega}\ vs\ 1/B_{ext}$ are plotted to separate thermal contributions $R_{\nabla T}$, $B_{DL}$ and $B_{FL+Oe}$ respectively.

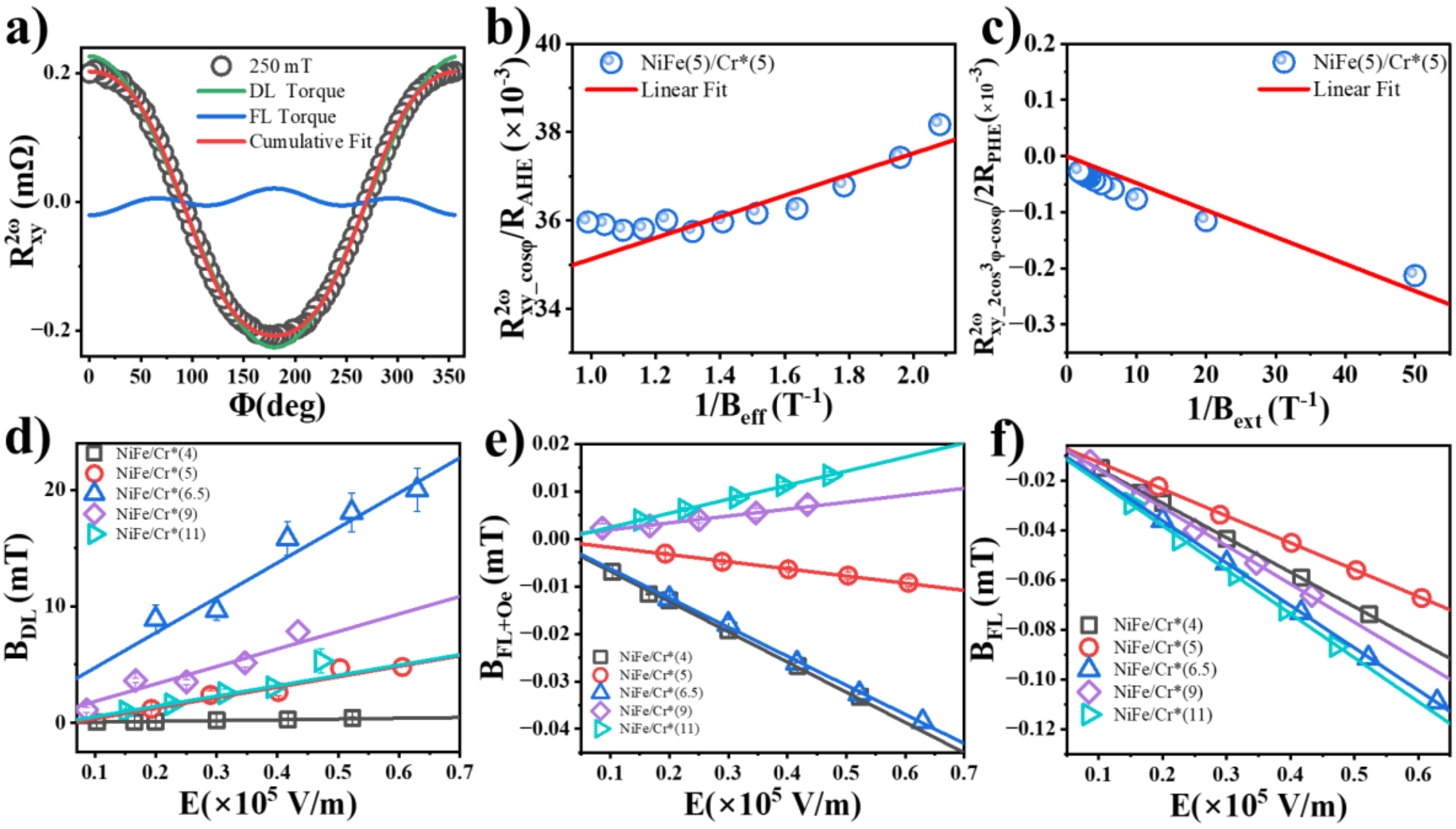

**Figure 4 |** (a) Second harmonic resistance of NiFe (5)/Cr* (5), fitted using equation (1) for 250 mT field at 3mA. (b), (c) depicts the linear fits of $R_{xy_\cos\varphi}^{2\omega}\ vs\ 1/B_{ext} + B_k$ and $R_{xy_2cos^3\varphi - cos\varphi}^{2\omega}\ vs\ 1/B_{ext}$ to extract $B_{DL}$ and $B_{FL+Oe}$. (d)Plot of damping like, (e) field like with Oersted field (effective field) (f) extracted field like torque from effective field normalized over Electric Field for different NiFe(5)/Cr*(t).

The corresponding linear fitting as depicted in Figure 4b,c yields the orbital torque induced effective fields for the stack NiFe(5)/Cr*(5). (See supplementary section S5 for the quantification of thermal contribution). A substantial amount of DL field $(2.39 \pm 0.3)$ mT with minimal $R_{\nabla T}$ $(0.21 \pm 0.003)$ mΩ and FL field $(0.033 \pm 0.004)$ mT was extracted for 5 nm thickness of Cr* device for $J_{Cr*} \sim 9.3 \times 10^9$ A/m$^2$. The observed field like torque is significantly smaller as compared to damping like torque. This procedure was systematically repeated across various Cr* thicknesses and applied electric fields to robustly quantify the torque generation (See S5).

To quantify the current-induced torque, the extracted effective fields are normalized to the applied electric field $E = IR^{\omega}_{xx}/l$, where $l$ is the length of the hall bar and $R^{\omega}_{xx}$ is the longitudinal resistance. It is evident from Figure 4d,e that both $B_{DL}$ and the effective field of $B_{FL+Oe}$ scale linearly with $E$ for different Cr* thickness, confirming its current-driven origin. Further, the $B_{FL}$ (Figure 4f) is estimated for different Cr* thickness by carefully quantifying and subsequently subtracting the corresponding Oersted field (See S6). Utilizing the slopes $(B_{DL,FL}/E)$ the torque efficiency is further substantiated as $\xi^E_{DL,FL} = \frac{2e}{\hbar} M_s t_{FM} \frac{B_{DL,FL}}{E}$ where $M_s$ is the saturation magnetization, $t_{FM}$ is the ferromagnetic layer thickness[2, 4, 29].

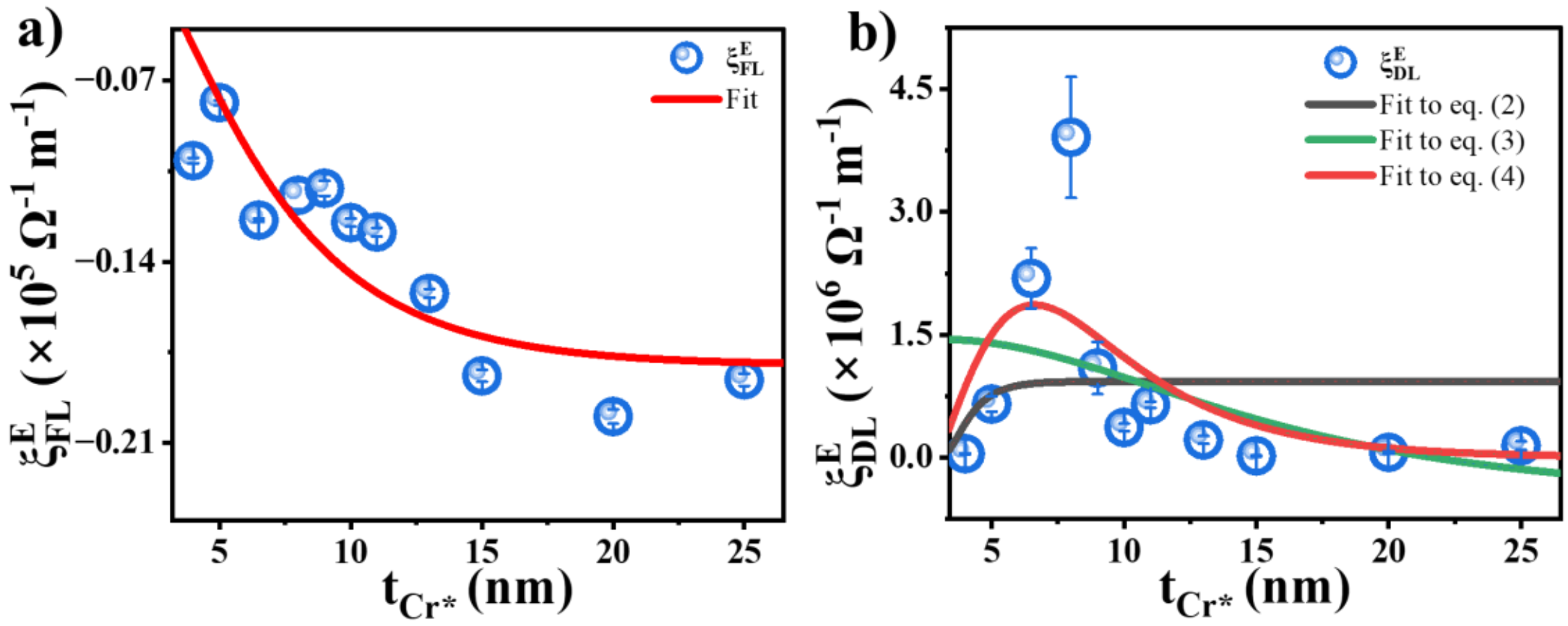

**Figure 5 |** (a) Field like efficiency fitted with equation 2 and (b) Damping like efficiency fitted through equation (2), (3) depicting the failure of conventional model (black line) and fit through proposed model, equation (4) (Red line).

Figure 5a,b depicts the variation of the $\xi^E_{FL,DL}$ with Cr* thickness. The effective damping like efficiency exhibits a pronounced non monotonic thickness dependence on Cr* thickness. It is only $(0.048 \pm 0.005) \times 10^6\ \Omega^{-1}\text{m}^{-1}$ for 4nm Cr* which increases to $(2.2 \pm 0.3) \times 10^6\ \Omega^{-1}\text{m}^{-1}$ for 6.5 nm and reaches a maximum of $(3.9 \pm 0.7) \times 10^6\ \Omega^{-1}\text{m}^{-1}$ for 8nm thick Cr* and decays to $(0.14 \pm 0.05) \times 10^6\ \Omega^{-1}\text{m}^{-1}$for 25 nm thickness of Cr* device. Notably, the peak value of $\xi^E_{DL}$ in our stack exceeds the conventional heavy-metal spin sources by one(two) orders of magnitude; for instance the highest observed DL efficiencies reported in the literature for different stacks are as follows: $\xi^E_{DL-Pt} = 2.43 \times 10^5\ \Omega^{-1}\text{m}^{-1}$ in Pt(5)/Co(1)/MgO [4], $\xi^E_{DL-Pt} = 2.33 \times 10^5\ \Omega^{-1}\text{m}^{-1}$ in Pt(4)/Co(3) [29] and $\xi^E_{DL-Ta} = -0.34 \times 10^5\ \Omega^{-1}\text{m}^{-1}$ in Ta(3)/CoFeB [47] , $\xi^E_{DL-Ta} = -0.12 \times 10^5\ \Omega^{-1}\text{m}^{-1}$ for Ta/PtCo/SiN[34]. Further, our values are comparable to the largest orbital-torque conductivities recently reported for the FM/LM-$O_x$ heterostructures based on Cu/$CuO_x$ ; such as $\sim 1.9 \times 10^6\ \Omega^{-1}\text{m}^{-1}$in crystalline $Cu_2O$/Cu/NiFe[48] ,$\xi^E_{DL}$ up to $\sim -1.48 \times 10^6\ \Omega^{-1}\text{m}^{-1}$in Cu*(7)/CoO(2)/Co(3) at low temp(20 K)[29] and $\sim 1.8 \times 10^6\ \Omega^{-1}\mathrm{m}^{-1}$ for Ta(3)/PtCo(3.3)/Cu(22)/SiN(3) [34]. Notably, achieving such a large value of $\xi_{DL}^{E}$ in our stack with a naturally oxidized 3d metal, without the need for a heavy-metal spin converter, is the central result of this work.

Remarkably, the field-like and damping-like torque efficiencies exhibit markedly different thickness dependences on Cr* thickness, and this contrast is itself significant. The variation of $\xi_{FL}^{E}$ is well described by the conventional diffusive orbital accumulation model of the form

$$\left(1 - \mathrm{sech}\left(\frac{t_{Cr^*}}{\lambda}\right)\right) \qquad (2)$$

which is expected for a single bulk-like source[4]. By Fitting the FL data, a characteristic transport length of $\lambda = 4.3 \pm 0.7$ nm can be estimated (Figure 5a). In contrast, the damping-like efficiency does not follow the same behaviour (Figure 5b).

In order to understand its thickness dependence, we solve the one-dimensional orbital drift-diffusion problem for the actual layer geometry. Our previous XPS measurements demonstrate that the native oxide is self-limiting, extending approximately ($t_{ox} \approx 3$ nm) from the free surface. Consequently, a film of nominal thickness (t) contains a metallic Cr channel of thickness ($d = t - t_{ox}$), bounded by NiFe at (z=0) and the Cr/CrO$_x$ interface at $z = d$. Here, the orbital accumulation satisfies the diffusion equation $\partial_z^2\, \delta\mu_L = \delta\mu_L/\lambda^2$, originating from two distinct sources where $\mu_L$ is the orbital accumulation and $\lambda$ is the effective diffusion length. The first source is the bulk orbital Hall effect in metallic Cr, which generates a uniform orbital current density ($j_B \propto \sigma_{OH} E$). The second is the orbital Rashba-Edelstein effect emerging at the inversion-asymmetric Cr/CrO$_x$ interface, which injects an interfacial orbital current $j_I(E)$ at $z = d$. Both orbital currents are converted into spin torque through the spin-orbit coupling in NiFe owing to its high Ni content, characterized by a common conversion efficiency $\eta$. The orbital current accumulated at the NiFe interface is therefore,

$$J(0) = j_B\,[1 - \mathrm{sech}(d/\lambda)] + j_I\,\mathrm{sech}(d/\lambda), \text{ such that}$$

$$\xi_{DL}^{E}(t) = \xi_B\left[1 - \mathrm{sech}\left(\frac{d}{\lambda}\right)\right] + \xi_I\,\mathrm{sech}\left(\frac{d}{\lambda}\right) \qquad (3)$$

Considering that both orbital-current sources are thickness independent, Equation (3) can be rewritten as $\xi_{DL}^{E}(t) = \xi_B + (\xi_I - \xi_B)\,\mathrm{sech}(d/\lambda)$ which is strictly monotonic with thickness. It simply interpolates between the interfacial limit at ($d \to 0$) and the bulk limit at ($d \to \infty$), and therefore cannot exhibit an interior maximum. This conclusion remains valid even when the interface is treated as non-ideal. For instance, introducing a finite orbital mixing conductance (G) at the NiFe interface gives $J(0) = \frac{\left[j_B\left(\cosh\left(\frac{d}{\lambda}\right)-1\right)+j_I\right]}{\left[\cosh\left(\frac{d}{\lambda}\right)+r\sinh\left(\frac{d}{\lambda}\right)\right]}$ with $r = \sigma_{Cr}/2eG$. The derivative of this expression is proportional to $(j_B - j_I)[\sinh(d/\lambda) + r\cosh(d/\lambda)] - j_B r$ which changes sign at most once, from negative to positive. Consequently, a constant-source two-channel model can produce at most a shallow interior minimum, but never a maximum (See detailed calculation in S7). Therefore, no combination of a thickness-independent bulk orbital Hall source and a thickness-independent interfacial orbital Rashba source, sharing the same transport length within the metallic channel, can reproduce the pronounced peak observed experimentally in Figure 5b. Furthermore, the fit based on equation (3), shown in Figure 5b, fails to account for the thickness dependence of $\xi_{DL}^{E}$ underscoring the inadequacy of this model to explain the experimental observations. The observed maximum thus requires the source term itself to evolve with Cr thickness, implying that the interfacial channel progressively develops as the film grows.

This interpretation is fully consistent with the oxidation chemistry revealed by XPS depth profiling. For stack with Cr* thickness 4 nm, the film is almost completely oxidized as the oxide layer thickness is 3nm approximated through XPS, leaving neither a continuous metallic Cr channel capable of transporting an injected orbital current nor a well-defined metal/oxide interface capable of generating one. We therefore introduce an oxidation-gated interfacial source, replacing ($\xi_I$) with ($\xi_I\, g(t)$), where $g(t)$ is activation function given by, $g(t) = \tanh\left(\frac{d}{\lambda}\right)$ which is diffusion alike while the bulk conductivity remains unchanged. The damping-like efficiency then becomes,

$$\xi_{DL}^{E}(t) = \xi_B \left[1 - \mathrm{sech}\left(\frac{d}{\lambda}\right)\right] + \xi_I\, g(t)\, \mathrm{sech}\left(\frac{d}{\lambda}\right) \quad (4)$$

Equation (4) possesses the generic drift-diffusion form of a rising source multiplied by a decaying transmission function in addition to the bulk term. This expression is similar to the well-known non-monotonic ferromagnet-thickness dependence of intrinsic current-induced torques[49]. Here, the $\tanh(d/\lambda)$ term describes the gradual build-up of the interfacial orbital-current source, whereas the $\mathrm{sech}(d/\lambda)$ term represents the transmission of that current through the metallic Cr channel to the NiFe layer. Moreover, the full model explicitly captures the finite bulk background contribution together with the gradual oxidation-induced activation of the interfacial source leading towards the orbital torque in NiFe layer.

We have used Equation (4) to fit the thickness dependence of $\xi_{DL}^{E}$ and the same is shown in Figure 5b, which quantitatively capture the observed non-monotonic behaviour. We have estimated an orbital transport length of $\lambda = (4.1 \pm 1.1)$ nm , consistent with the value independently obtained from the field-like torque $(4.3 \pm 0.7)$ nm fit. The close agreement between the transport lengths extracted independently from the damping-like and field-like torques supports the validity of the proposed orbital transport model. The extracted values are smaller than the ($6.6 \pm 0.6$ nm) nm orbital transport length measured magneto-optically in epitaxial Cr and Cr based heterostructure[17, 40]. This comparison suggests that native oxidation, while creating the dominant interfacial torque channel, simultaneously introduces additional disorder that limits orbital-current propagation.

The interfacial and bulk contribution is approximately found to be $\xi_I \sim (3.7 \pm 1.02) \times 10^6\ \Omega^{-1}\mathrm{m}^{-1}$ and $\xi_B \sim (0.29 \pm 0.12) \times 10^6\ \Omega^{-1}\mathrm{m}^{-1}$ by fitting the equation (4) of non-monotonic dependence of $\xi_{DL}^{E}$ on Cr* thickness. The interfacial contribution is found to be larger by approximately 13 times than the bulk orbital Hall channel. Furthermore, the extracted interfacial coefficient is nearly one order of magnitude larger than both the experimentally measured ($5 \times 10^5\ \Omega^{-1}\mathrm{m}^{-1}$) and theoretically predicted ($4 \times 10^5\ \Omega^{-1}\mathrm{m}^{-1}$) bulk orbital hall conductivity of Cr[11, 50]. Together with the comparatively small bulk contribution obtained from the two-channel analysis, this independently confirms that the giant damping-like torque cannot originate from the bulk orbital Hall effect alone. This interpretation is fully consistent with our first-principles calculations, which reveal that oxygenation enhances the orbital Hall conductivity of the NiFe/Cr heterostructure by nearly a factor of three through Cr(3d)-O(2p) hybridization localized at the oxidized interface. More broadly, these findings establish an important design principle for orbital-torque materials: the native oxide interface is not merely a perturbation to the bulk orbital Hall channel; rather, it fundamentally redefines the origin of orbital torque by transforming a weak bulk source into a dominant interfacial generator of orbital current.

This interface-dominated mechanism motivates us to carry out two clear and experimentally testable predictions. First, suppressing the native oxide should strongly reduce the torque even though the metallic Cr layer, thus the bulk orbital Hall effect, remains intact. Second, interrupting the direct NiFe/Cr* interface while preserving the oxide should also strongly suppress the torque. We tested both predictions using control samples (See S8).

In the capped control sample of NiFe(5)/Cr(6.5)/Ta(2), the 2 nm Ta cap prevents the natural oxidation of Cr, thereby eliminating the Cr/$CrO_x$ interface while leaving the metallic Cr layer unchanged. Notably, the buried NiFe/Cr interface remains the same in both the capped and uncapped samples. Thus, the orbital to spin conversion process is expected to remain unchanged in both the cases. Therefore, any difference in torque directly reflects the role of orbital current generation at the oxidized surface. The extracted damping-like field for the capped sample is only $0.035 \pm 0.01\ mT$, which is more than two orders of magnitude smaller than the $8.91 \pm 1.23\ mT$ measured for the naturally oxidized NiFe(5)/Cr*(6.5) device at the same current density, $J_{Cr^*} \approx 5 \times 10^9$ A/m$^2$ (Figure S8(c)). The significant reduction therefore shows that these oxidation-independent mechanisms emanating from the bulk Cr channel cannot explain the giant torque. This result independently corroborates that the bulk contribution $\xi_B$ estimated from the drift-diffusion analysis is only a minor component.

To test the second hypothesis, we have prepared NiFe(5)/Cu(1)/Cr*(6.5) stack where the Cr layer undergoes natural oxidation just as in the main samples, so the interfacial orbital-current source is preserved. However, a 1-nm Cu spacer interrupts the direct NiFe/Cr* interface. In this case, the damping-like field estimated to ~ $0.077 \pm 0.02\ mT$, representing more than a hundred-fold reduction despite the presence of the oxidized Cr surface. Such a large decrease cannot be explained simply by attenuation during transport through a 1-nm Cu layer. Instead, it indicates that efficient orbital-current injection requires direct orbital hybridization at the NiFe/Cr* interface. The Cu spacer, whose electronic states near the Fermi level are predominantly of $s$ character, weakens this hybridization and strongly reduces the transfer of orbital angular momentum into NiFe, where it is converted into spin[51].

Together, these two control experiments clearly separate the roles of orbital-current generation and conversion. Removing the Cr/$CrO_x$ interface suppresses the orbital-current source, whereas inserting the Cu spacer suppresses the efficient conversion of orbital current at the NiFe interface. Both processes are therefore essential for producing the giant damping-like torque, providing direct experimental evidence that the torque originates at the oxidized Cr surface, propagates through the metallic Cr layer, and is converted into spin at the NiFe interface.

To further investigate the microscopic origin of the current-induced torque, we measured the orbital unidirectional magnetoresistance (OUMR) in the NiFe/Cr* devices. The OUMR is a nonreciprocal resistance which scales linearly with the applied current and transverse magnetization component as $j[\mathbf{m} \cdot (\hat{z} \times \hat{j})] = j\, m_y$. Further, it reverses its sign upon reversing either the current or the magnetization[52-53]. Notably, it emerges as the second-harmonic contribution to the longitudinal resistance ($R_{xx}^{2\omega}$).

Microscopically, the UMR originates from the nonequilibrium spin or orbital accumulation generated at the interface with the ferromagnet, leading to a resistance that depends on the relative orientation between the accumulated angular momentum and the magnetization[54-56]. In the case of orbital currents, this modulation arises from orbital dependent electron scattering, such as orbital selective *s–d* transitions together with orbital to spin conversion inside the ferromagnet. An orbital analogue of the UMR has been identified in naturally oxidized Cu*/Co bilayers[53] and in Ni/Ti and Ni/Nb bilayers where it scales directly with the orbital torque efficiency[52]. Since the same nonequilibrium orbital accumulation is expected to drive both the damping-like torque and the Orbital UMR, the OUMR thus provides an independent transport probe of orbital current generation that does not rely on harmonic Hall torque measurements.

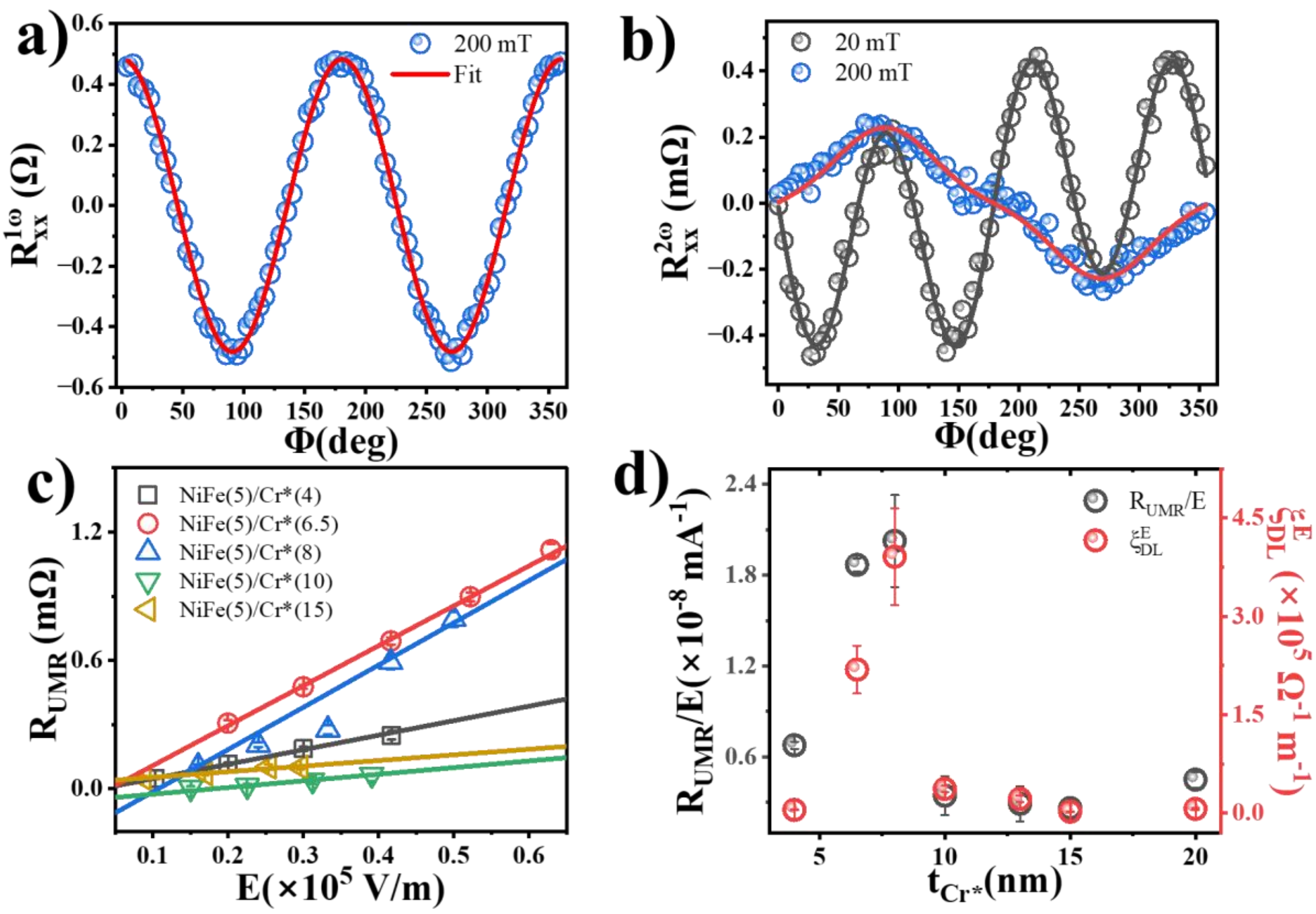

**Figure 6 |** (a) First and (b) second harmonic angular magnetoresistance of NiFe(5)/Cr*(4), solid lines represent fit obtained from equation 5 and 6. (c) Orbital UMR normalized over applied electric field for various NiFe(5)/Cr*(t), straight lines are the linear fits used to obtain $R_{UMR}$ per electric field. (d) Correlation of damping like torque efficiency with the $R_{UMR}$ for different thickness of NiFe/Cr*.

Figure 6a,b shows the first and second harmonic longitudinal resistances for NiFe(5)/Cr*(4) at 2mA current which are fitted using equation (5) and (6) respectively,

$$R_{xx}^{1\omega} = \Delta R_{xx}^{1\omega}\cos^2\varphi \tag{5}$$

$$R_{xx}^{2\omega}(\varphi) = R^*\sin\varphi - 2\Delta R_{xx}^{1\omega}\frac{B_{FL+Oe}}{B_{ext}}(\cos^2\varphi \cdot \sin\varphi) \tag{6}$$

where the coefficient $R^*$ carries the characteristic $m_y \sim \sin\varphi$ symmetry of the OUMR[53-54]. It contains both the orbital UMR and a transverse thermal contribution, $R^* = R_{UMR} + hR_{\nabla T}$ where the geometric factor $h = \Delta R_{xx}^{1\omega}/\Delta R_{xy}^{1\omega} \sim 1.1$ agrees well with the lithographic aspect ratio $(l/w)$ of the Hall bars and is largely insensitive to patterning imperfections. The thermal contribution is removed using $R_{\nabla T}$ extracted independently from the second-harmonic Hall measurements (See S5). The angular dependence exhibits four extrema at the lower magnetic field (20 mT) (Figure 6b), indicating a significant contribution from the effective field $B_{FL+Oe}$ which gets suppressed at higher fields. At sufficiently large magnetic fields which suppresses the magnon contribution, orbital $R_{UMR}$ scales linearly with the applied electric field (Figure 6c), as expected. Generally, the low-field behaviour of the UMR provides a sensitive probe for distinguishing orbital and spin transport. In conventional FM/HM bilayers, spin currents excite or annihilate magnons in the ferromagnet, enhancing electron-magnon scattering and producing a pronounced increase in the UMR at low magnetic fields. Since magnons carry spin but not orbital angular momentum, such an enhancement in light-metal systems indicates the presence of a spin-current component, originating either from a residual spin Hall effect in the nonmagnetic layer or from orbital-to-spin conversion within the ferromagnet[52-53].

Consistent with a predominantly orbital-current mechanism, NiFe/Cr* exhibits essentially no low-field enhancement for thicknesses such as 4 and 20 nm, while only a very weak enhancement appears for 6.5 and 8 nm, coinciding with the maximum of $\xi_{DL}^{E}$ (See S9). This thickness correlation indicates that the magnon contribution arises primarily from orbital-to-spin conversion inside NiFe, since the converted spin current is proportional to the orbital current arriving at the interface and therefore peaks together with the torque. In contrast, a spin Hall current generated directly in Cr would increase and eventually saturate with Cr thickness, and should therefore remain detectable at 20 nm, contrary to experiment. This interpretation is consistent with first-principles calculations predicting the spin Hall conductivity of Cr to be more than an order of magnitude smaller than its orbital Hall conductivity, as well as recent experiments identifying metallic Cr as an efficient orbital-current source with negligible spin Hall contribution[57].

We further correlate the damping-like efficiency with the orbital UMR normalized per unit electric field, $R_{UMR}/E$, as a function of Cr* thickness (Figure 6d) Remarkably, both exhibits nearly identical non-monotonic thickness dependence, increasing to a maximum near 8 nm before decreasing for thicker Cr* layers. Such correlated scaling between the orbital UMR and the damping-like torque has previously been established as a hallmark of orbital current driven transport in Ni/Ti, Ni/Nb[52],Cu*/Co[53] and Co-O/Pt/Co [22] heterostructures indicating that in our stack nonequilibrium orbital current is generated within the Cr* layer.

One has to note that the damping-like torque and the orbital UMR probe different physical processes, though they are driven by the same nonequilibrium orbital accumulation arriving at the NiFe interface. The torque depends on the transmission of orbital current into NiFe followed by orbital to spin conversion, whereas the orbital UMR primarily probes the interfacial orbital accumulation through orbital dependent scattering and is expected to be comparatively less sensitive to interfacial transmission. Their nearly identical thickness dependence therefore indicates that the dominant thickness evolution is governed by the orbital current source itself rather than by changes in the orbital to spin conversion efficiency of NiFe. At small Cr thicknesses, the interfacial OREE channel is suppressed because the Cr layer is almost completely oxidized and neither a well-defined $Cr/CrO_x$ interface nor a continuous metallic transport channel exists. As the metallic Cr thickness increases, the interfacial source becomes fully established, giving rise to the rapid increase in both the torque and the orbital UMR. Beyond the optimum thickness, however, the orbital current generated at the remote $Cr/CrO_x$ interface is progressively attenuated during diffusion through the metallic Cr layer, while current shunting in the thicker Cr further reduces the effective driving field. Consequently, the interfacial contribution decreases, leaving the slowly saturating bulk orbital Hall channel to dominate the response.

Taken together, the first-principles calculations, the drift–diffusion analysis, and the independent orbital UMR measurements consistently support a two-channel mechanism where a conventional bulk orbital Hall current coexists with a substantially stronger interfacial orbital Rashba–Edelstein current whose magnitude is controlled by the oxidation profile of the Cr layer, for orbital-current generation in NiFe/Cr*. The orbital UMR thus provides an independent transport signature of the same orbital current responsible for the damping-like torque, offering direct experimental support for the microscopic mechanism proposed in this work.

Next, we demonstrate current-induced magnetization switching in the deposited NiFe/Cr* stack. Generally, the critical switching current density scales inversely with the damping-like torque efficiency for a fixed ferromagnet[58-60], the giant $\xi_{DL}^{E}$ observed in our sample should translate directly into reduced switching currents. We first examine the magnetic anisotropy of the device through low-field planar Hall measurements for NiFe/Cr*(6.5) (Figure 7a), which reveal that the in-plane easy axis is not collinear with the current direction but carries a built-in tilt. We indeed observe the field-free current-induced switching across the NiFe/Cr*(t) series (See S10).

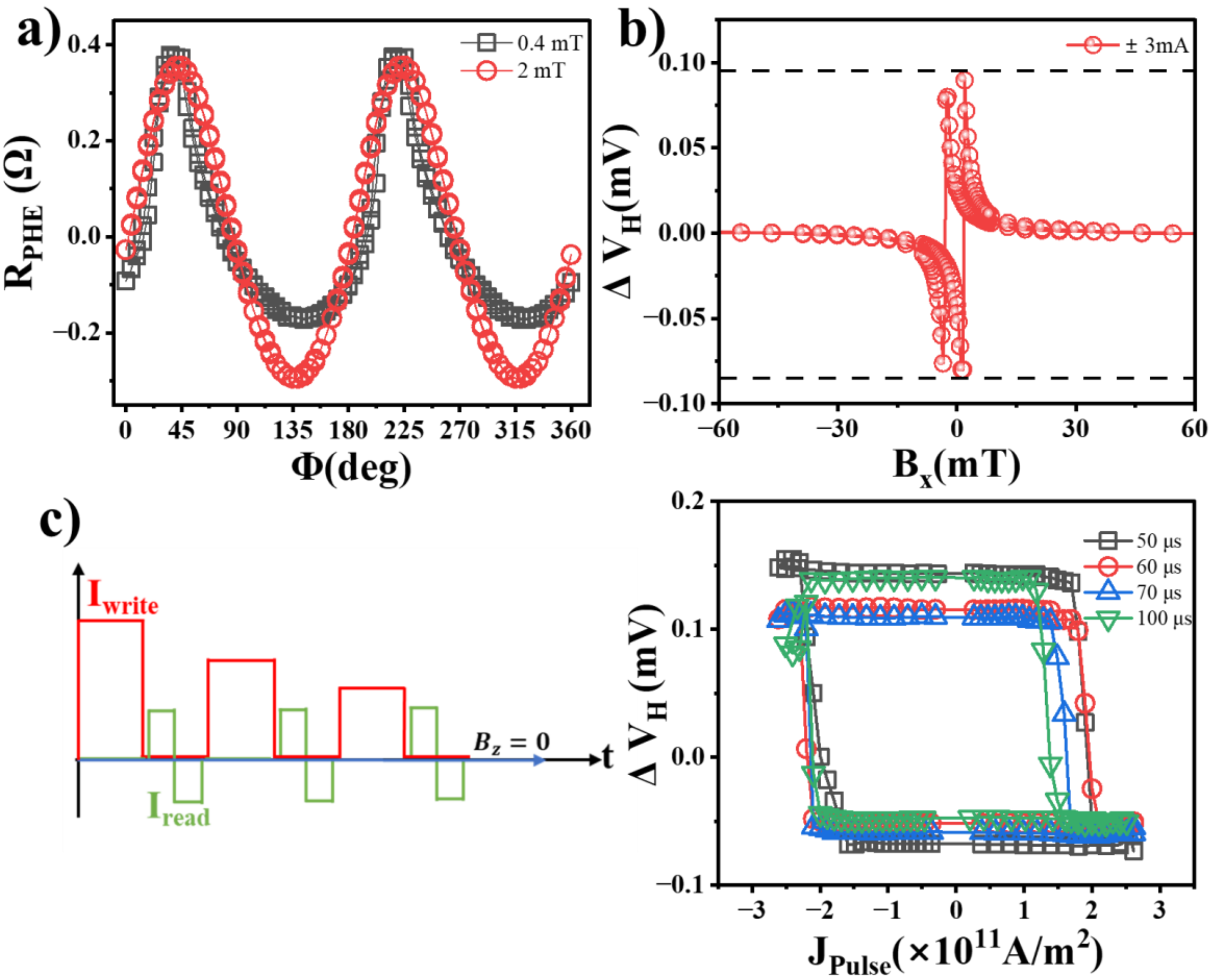

**Figure 7 |** (a) Low field PHE to detect the type of device.(b) Reading of magnetic states by using OPHV method where Δ $V_H$ is separated by dashed line. (c) Schematic describing the method used to perform Field free switching measurement where after the write pulse application a read current ($\pm I$) is applied and corresponding output for Nife(5)/Cr*(6.5) device at different pulse widths.

The magnetic switching characteristics for NiFe/Cr*(6.5) device were investigated using the odd planar Hall voltage technique[61] (Figure 7b), which clearly resolve the two magnetic states separated approximately by $\Delta\, V_H = 0.17$ mV. The method for the switching experiment is depicted in Figure 7c, where the magnetic state is detected after each current pulse by a small read current of $\pm 3$ mA at zero field. We have taken the average of 10 readings to obtain the magnetic state corresponding to a particular writing current pulse for better signal to noise ratio. Figure 7c shows field-free switching of the NiFe(5)/Cr*(6.5) device for different write-pulse widths. The close agreement between the readout amplitudes obtained from the OPHV (Figure 7b) and current induced field free switching (Figure 7c) measurements confirms complete and reliable magnetization reversal. The field-free switching current density for this stack is $1.58 \times 10^{11}\ A/m^2$ at $50\mu s$ pulse width.

To isolate the role of oxidation, identical measurements were performed on the control sample Cr (6.5)/NiFe (5), in which Cr layer is deposited beneath the NiFe layer and is thereby protected from oxidation. No significant field-free switching is observed in this device. We have applied an out-of-plane symmetry-breaking field of 0.5 mT to elucidate the current induced switching in this control device. Notably even after applying an out of plane field the switching current density, $2.2 \times 10^{11}\ \mathrm{A/m^2}$ exceeds the field-free value of the oxidized device by ≈ 40%, while the switching transitions are markedly less abrupt, indicating a substantially weaker torque (Figure S10.2(b)).

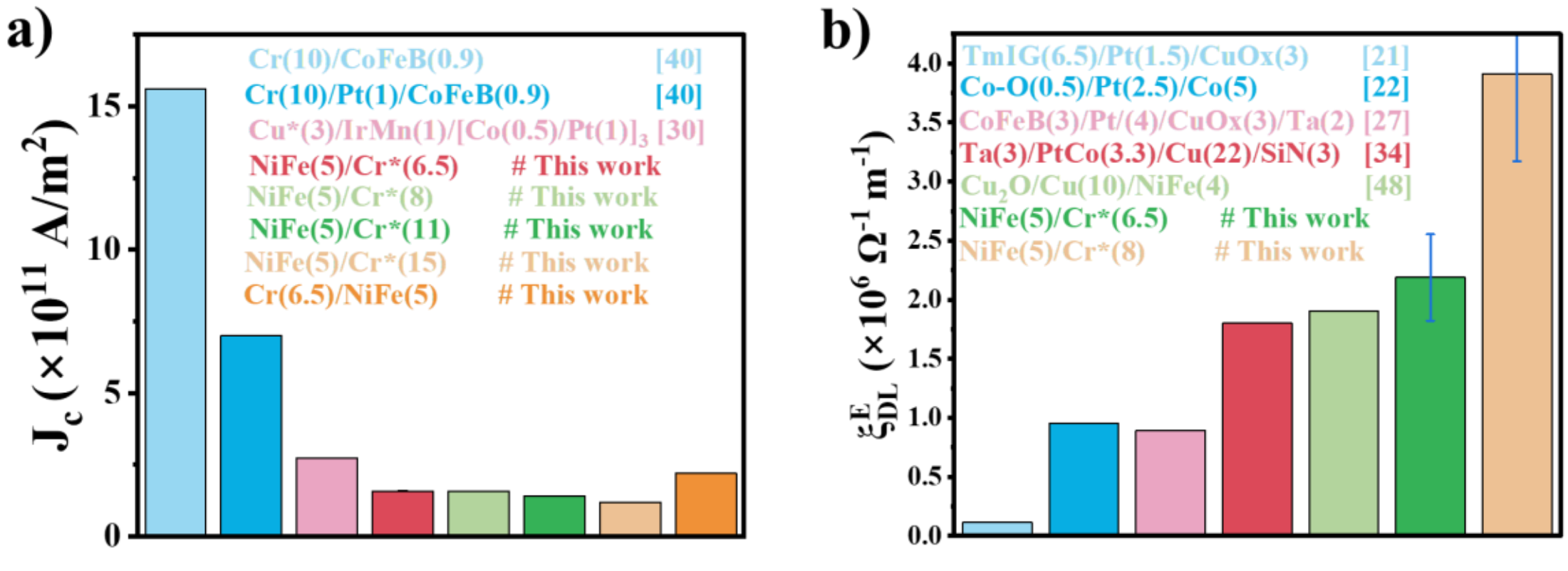

**Figure 8 |** Comparison of (a) Switching current density where external magnetic field assistance was required by Ref.[30, 40] whereas present work does not require any external field assistance and (b) large damping like efficiency of the present work(NiFe(5)/Cr*(t)), with other works based on light metal or light metal oxide orbital current sources.

The switching performance compares favourably with previously reported orbital torque systems in Figure 8a (See S11). The field-free $j_c$ of NiFe(5)/Cr*(6.5) is roughly an order of magnitude lower than that of Cr(10)/CoFeB(0.9) ($1.5 \times 10^{12}\ A/m^2$) and more than fourfold lower than Cr(10)/Pt(1)/CoFeB(0.9) ($0.7 \times 10^{12}\ A/m^2$), both of which additionally require a 20 mT symmetry-breaking field[40]. The observed switching current densities in our stack is also lower than those reported for conventional heavy-metal spin current devices such as Pt(3)/CoFeB(2) and Ta(3)/CoFeB(2), $3.3 \times 10^{11}\ A/m^2$ and $2.2 \times 10^{11}\ A/m^2$ ,respectively[25]. Our value is furthermore below than the field-free type-Y switching of the heavy-element-free CoFeB(7)/$CuO_x$(4) device[25] at current density of $2.7 \times 10^{11}$ A/$m^2$ and comparable to TmIG(6.5)/Pt(1)/CuOx(3) ($1.7 \times 10^{11}$ A/$m^2$), which additionally requires a 2.5 mT assist field and a Pt conversion layer[21]. The NiFe/Cr* heterostructure achieves these figures without heavy metals, external fields, or additional spacer and oxide engineering; the functional interface is created by native oxidation alone. These comparisons establish interfacial Cr oxidation as an efficient route for field free orbital-torque-driven magnetization switching.

## 3| Conclusion

Figure 8a,b compares the switching and damping like efficiency of present NiFe/Cr* system with recently reported other orbital torques based systems[21-22, 27, 30, 34, 40, 48]. In conclusion, we have demonstrated that the self-limiting native $CrO_x$ formed on Cr in NiFe/Cr bilayers provides a simple and efficient platform for generating giant orbital torque. We have obtained a damping-like efficiency, $\xi_{DL}^{E}$, of $(3.9 \pm 0.7) \times 10^6\, \Omega^{-1}m^{-1}$ at optimum Cr* thickness which is one to two orders of magnitude larger than those of conventional Pt- and Ta-based spin current sources and exceeds the highest efficiencies reported for engineered $CuO_x$ systems. We have achieved deterministic field-free magnetization switching at a current density of only $1.58 \times 10^{11}$ A/$m^2$. Importantly, the native oxide not only generates the orbital current but also induces the in-plane anisotropy tilt required for field-free switching. As a result, orbital-current generation, torque conversion, and deterministic switching are realized in a simple NiFe/Cr* bilayer without heavy metals, external symmetry breaking magnetic fields, spacer layers, or deliberately engineered oxide interfaces.

Our combined experimental and theoretical analyses elucidate the microscopic origin of this phenomenon. First-principles calculations show that oxygenation enhances Cr(3d)–O(2p) hybridization, leading to an approximately threefold increase in the orbital Hall conductivity of the NiFe/Cr heterostructure. Furthermore, a quantitative drift-diffusion analysis further shows that the observed thickness dependence cannot be explained by bulk orbital Hall contribution alone. Instead, it requires a dominant interfacial orbital Rashba–Edelstein source that is activated by oxidation and transmitted through the metallic Cr layer over a characteristic length of approximately 4 nm, together with a much smaller bulk orbital Hall contribution. This interpretation is independently supported by two control experiments; suppressing the native oxide or interrupting the direct NiFe/Cr* interface reduces the torque by more than two orders of magnitude. Furthermore, the close correspondence between the orbital unidirectional magnetoresistance and the damping like torque efficiency across the entire thickness series provides independent transport evidence that both originate from the same orbital source.

Beyond establishing naturally oxidized Cr as an efficient orbital-torque source, our work shows that native oxidation of Cr should not be viewed merely as a parasitic surface effect, but rather as a functional mechanism that creates the interface responsible for giant orbital-current generation. Since the naturally formed oxide under ambient conditions is self-limiting, this strategy requires no additional fabrication steps or heavy-metal conversion layers. Broadly, our work highlights oxidation-controlled interfaces as a promising route for designing efficient orbital-current sources and suggests that similar mechanisms may be realized in other light metals that form stable native oxides, thereby expanding the materials platform for low-power orbitronic memory and logic technologies.

## 4| Experimental Methods

The stacks were grown by DC magnetron sputtering in a high-vacuum chamber with a base pressure of $8.7 \times 10^{-8}$ mbar. The 2 nm Ta seed layer was used to promote adhesion, and the Cr* layer on top of NiFe serves as the orbital-current source. Additional control samples, including a Ta-capped sample (Ta(2)/NiFe(5)/Cr(6.5)/Ta(2)) and a sample with a different interface Ta(2)/NiFe(5)/Cu(1)/Cr*(6.5)), were prepared to isolate the respective contributions of the oxidized Cr and the Cr/NiFe interface. To obtain an oxygen gradient in the Cr layer and utilize it as an interfacial orbital source, the devices were allowed to naturally oxidize in ambient condition over a period of 48 hours prior to measurement. The root-mean-square surface roughness of the resulting heterostructures, measured by atomic force microscopy, is exceptionally smooth at approximately 0.42 nm (See S2).

Magnetization data was collected for blanket layers using Physical property measurement system (PPMS, Quantum design)-VSM which shows the in plane easy axis of the samples (See S4). Depth profiling XPS was carried out using a Thermofisher Escalab Xi+ system with Ar+ ion etching to identify the chemical state of Cr, O, Ni, Fe in the NiFe(5)/Cr*(6.5) and NiFe(5)/Cr*(11). The obtained peaks were fitted using the Avantage (©) software.

The deposited films were patterned into 15 × 15 μm² Hall bars by maskless photolithography and lift-off, followed by a second lithography step to define Ta(5)/Ag(50) contact electrodes. Subsequently, the samples were wire-bonded for transport measurements. AC current at 577.13 Hz of varying amplitude(1-6mA) was sourced through the device with a Keithley 6221, and the first ($R_{xy}^{1\omega}$)and second ($R_{xy}^{2\omega}$) harmonic transverse voltages were measured with an EG&G 7265 lock-in amplifier while rotating the sample in presence of an external in-plane field. All the measurements are performed at room temperature.

**Acknowledgements**

CK acknowledges Raghavendra Posti for the insightful discussions and CRF IIT Ropar for research facilities. DR acknowledges project no. ANRF/ARG/2025/009882/PS. This work is supported by the U.S. Department of Energy, Office of Basic Energy Sciences, Division of Materials Sciences and Engineering under Award No. DE-FG02-96ER45579. MKM and PJ extend their acknowledgment to the High-Performance Research Computing (HPRC) core facility at Virginia Commonwealth University for providing supercomputing resources. MKM acknowledges the ANRF Ramanujan Faculty fellowship under Award No. RJF/2025/000601. The infrastructure and research facility provided by IIT-Bhubaneswar are gratefully acknowledged.

**Conflicts of Interest**

The authors declare no conflicts of interest.

**Data Availability Statement**

The data that support the findings of this study are available upon reasonable request from corresponding author.